# Orbital order-disorder transition in doped perovskite manganites: influence of intrinsic octahedral-site distortion

Parthasarathi Mondal, Dipten Bhattacharya,[*] and Pranab Choudhury

*Nanostructured Materials Division, Central Glass and Ceramic Research Institute, CSIR, Kolkata 700 032, India*

The orbital order-disorder transition temperature ($T_{OO}$) versus tolerance factor ($t$) plot switches from monotonic to non-monotonic beyond a doping level ($x$) ~10 atom% in a family of $R_{1-x}A_x MnO_3$ systems (R = La, Pr, Nd; A = Ca, Sr; x = 0.0-0.2). $T_{OO}$ reaches maximum at a 'doping-dependent' critical tolerance factor $t_C(x)$ (for $0.1 < x \leq 0.2$) at which the orthorhombic distortion ($D$) also maximizes. Such an observation reflects influence of charge carriers on both 'intrinsic' octahedral-site distortion itself and its bias on the orbital order in doped perovskite manganites and, thus, deviation from what has been observed in undoped $RMnO_3$, $RTiO_3$, and $RVO_3$ systems where maximization of $T_{OO}$ and $D$ takes place at a universal tolerance factor or R-site ion size.

The orbital order in perovskite $RMnO_3$ is driven by (i) strong electron-electron superexchange interaction resulting from $Mn^{3+}$ ($d^4$) $e_g^1$ level degeneracy[1] (purely electronic), (ii) cooperative Jahn-Teller effect which results in lifting of the degeneracy and splitting of the Mn-O bonds[2] (electron and lattice), and (iii) enhanced R-O bond covalency[3] (electron and lattice). Although, it has been shown[4] that superexchange interaction alone is strong enough to give rise to as high an orbital order-disorder transition point $T_{OO}$ as ~650 K, large $T_{OO}$ above ~650 K and progressive increase in $T_{OO}$ with the decrease in average R-site radius $<r_R>$[5] or tolerance factor $t$ $[=(R-O)/\sqrt{2}(Mn-O)]$ indicates an important role of the electron-lattice interaction. Given this backdrop, it is tempting to check how strong is the competition between the 'electron-electron' and 'electron-lattice' interaction mechanisms in driving the long-range orbital order and which mechanism plays the dominant role in a variety of orthorhombic perovskite oxides such as $RMnO_3$, $RTiO_3$, $RVO_3$ etc together with doped systems such as $R_{1-x}A_xMnO_3$. Even though, smaller $R^{3+}$ ion was traditionally considered to give rise to only tilt and rotation of the $MO_6$ (M = Ti, V, Fe, Mn,…) octahedra around the crystallographic [110] and [001] axes, in a few recent works,[6] Goodenough and his coworker have shown that it yields an 'intrinsic' orthorhombic distortion as well. The $MO_6$ octahedra do not remain rigid. Interestingly, this 'intrinsic' orthorhombic distortion is shown[6] to be reaching a maximum at an intermediate average R-site radius $<r_R>$ ~1.110 Å if the <M-O> bond length is about 2.0 Å. $T_{OO}$ is found to be strongly governed by this 'intrinsic' distortion and thus exhibits a peak at $<r_R>$ ~1.110 Å in undoped $RMnO_3$, $RVO_3$ and $RTiO_3$. We explore, in this paper, what happens in the doped perovskite managnite series $R_{1-x}A_xMnO_3$ (R = La, Pr, Nd; A = Ca, Sr; x = 0.0-0.2) – whether the 'intrinsic' orthorhombic distortion biases $T_{OO}$ strongly or the correlation remains weak. The results could highlight the dominant mechanism behind the orbital order – electron-electron superexchange or electron-lattice interaction. While the finite charge carrier concentration promotes double exchange, both electron-electron superexchange and electron-lattice interaction weaken due to decrease in Jahn-Teller $Mn^{3+}$ ion concentration. The R-O bond covalency, on the other hand, enhances because of the presence of smaller $R^{3+}$ and $Mn^{4+}$ ions in the perovskite lattice. Therefore, it is important to investigate the orbital order-disorder transition in the entire family of rare-earth perovskite doped manganites, especially within a low doping regime $0.0 \leq x \leq 0.2$, to find out how the transition point $T_{OO}$ varies when both $x$ and $t$ are varied simultaneously.

In this paper, we show that the $T_{OO}$ versus $t$ pattern switches from monotonic to non-monotonic beyond $x \sim 0.1$ in the $R_{1-x}A_xMnO_3$ (R = La, Pr, Nd; A = Ca, Sr; x = 0.0-0.2) family. The non-monotonic $T_{OO}$-$t$ pattern in doped $RMnO_3$ replicates the trend observed[7] in undoped $RMnO_3$, $RTiO_3$, $RVO_3$ etc and, thus, indicates maximization of 'intrinsic' octahedral-site distortion of the lattice *albeit* at a 'doping-dependent' critical tolerance factor $t_C(x)$. We also explore how the rate of variation of $T_{OO}$ with $t$ – i.e., the slope $\Delta T_{OO}/\Delta t$ – depends on the doping level $x$. This plot will highlight the impact of overall octahedral tilt and distortion on $T_{OO}$ as a function of charge carrier concentration.

The experiments were, primarily, carried out on high quality phase pure bulk polycrystalline sintered pellets of $R_{1-x}A_xMnO_3$ (R = La, Pr, Nd; A = Ca, Sr; x = 0.0-0.2) compounds. In addition, single crystals of following compositions were also used to verify the results obtained in polycrystalline samples: $La_{0.95}Sr_{0.05}MnO_3$, $La_{0.89}Sr_{0.11}MnO_3$, $La_{0.95}Ca_{0.05}MnO_3$, $La_{0.9}Ca_{0.1}MnO_3$. The details of the synthesis of bulk polycrystalline pellets and single crystals are available in our earlier papers.[8,9] The orbital order-disorder transition was studied by dc resistivity and calorimetry measurements. The room temperature x-ray diffraction patterns were used to evaluate the lattice parameters, orthorhombic distortion, and other structural parameters for all the compounds.

In Fig. 1a, we show the representative dc resistivity ($\rho$) versus temperature ($T$) patterns for doped manganites. The transition point $T_{OO}$ is evaluated from the anomalous peak in the $d\ln(\rho/T)/d(1/T)$ versus $T$ plots (Fig. 1b). The width of the peak indicates the transition zone. The zone widens with the increase in doping level '$x$'. Below and above the zone, the $\rho$-$T$ pattern could be fitted with adiabatic small polaron hopping transport model $\rho = \rho_0 T^{\alpha}.\exp(E_A/k_BT)$ ($\alpha = 1.0$). The fitting yields the activation energy $E_A$ varying within ~0.1-0.5 eV. The widening of the transition zone has got a correlation with the latent heat ($L$) of transition measured from calorimetric studies – $L$ decreases sharply with the increase in '$x$' in $La_{1-x}A_xMnO_3$ (A = Ca, Sr) series (Fig. 1c) and reaches almost zero at $x = 0.05$. The transition zone – identified from $\rho$-$T$ data – also widens with the increase in '$x$'. In $Pr_{1-x}A_xMnO_3$ and $Nd_{1-x}A_xMnO_3$ series, of course, latent heat was found to be zero in global calorimetry across the entire range of '$x$'.

Figures 2a,b show the variation of lattice parameters $a$, $b$, $c/\sqrt{2}$ and lattice volume $V$ as a function of '$x$' – estimated from the room temperature x-ray diffraction data – for all the compounds. In all the cases the relation $c/\sqrt{2}<a<b$ – characteristic of O'

orthorhombic phase – is maintained. And finally, in Fig. 3a, we show the variation in $T_{OO}$ with $t$ for x = 0.0, 0.025, 0.05, 0.1, 0.125, 0.15, and 0.2. While the $T_{OO}$-$t$ is monotonic with rise in $T_{OO}$ for drop in $t$ in the low-doping regime ($x \leq 0.1$), it switches to non-monotonic with the appearance of a peak at a critical $t_C(x)$ in the high-doping regime ($0.1 < x \leq 0.2$). The switch in the $T_{OO}$-$t$ patterns and the 'doping-dependent' $t_C(x)$ are the *central results* of this paper.

The 'intrinsic' octahedral-site distortion in the perovskite lattice is measured by the split in M-O bond lengths (in-plane and across the plane) described by the degenerate vibration modes $Q_2$ [= $\ell_x - \ell_y$] and $Q_3$ [= $(2\ell_z-\ell_x-\ell_y)/\sqrt{3}$] where $\ell$ is the displacement originated from the split. It is also measured by the deviation of O-M-O angle ($\alpha$) from 90°, $D$, and lattice parameter $b$.[6,10] The tilt in the octahedra, on the other hand, is quantified by the M-O-M angle $\Phi = 180° - \omega$ and the Goldschmidt tolerance factor $t$. By analyzing the structural data, it has been shown[6] that for both $RFeO_3$ – with no orbital order – and $RVO_3$ – with $t_{2g}$ orbital order – the split in the M-O bond lengths, the average <M-O> bong length, and deviation of $\alpha$ from 90° maximize at a universal intermediate average R-site radius $<r_R>_U$ ~1.110 Å. It has also been shown[6] that $T_{OO}$ too, in $RVO_3$ series, reaches maximum at ~1.110 Å. The $GdFeO_3$ tilt and rotation of the octahedron, on the other hand, gives rise to progressively reduced average M-O-M bond angle $<\cos^2\Phi>$ with the decrease in $<r_R>$ and $t$. In other words, while $t$ and $\omega$ change monotonically with the increase in R-site ion size, parameters, such as, $\alpha$, split in the M-O bond lengths and average <M-O> bond length exhibit anomalous features around $<r_R>_U$. This *anomaly* points out that the crossover from orthorhombic phase to cubic is not smooth with the increase in $t$. An intermediate phase appears because of the influence of 'intrinsic' octahedral-site distortion. The extent of 'anomaly' is the measure of strength of the 'intrinsic' octahedral-site distortion associated with tilt and rotation of the $MO_6$ octahedron. $T_{OO}$ is governed more by this 'intrinsic' octahedral-site distortion than by any other factor in almost all the orthorhombic perovskite systems which exhibit long-range orbital order. Interestingly, doped manganite series does depict maximization of $D$ and $T_{OO}$ only at a doping dependent $t_C(x)$ (Fig. 3). *Unlike undoped systems, there is no universal tolerance factor in this case.* We plot $D$, the lattice parameter '$b$', and the average <Mn-O> bond length as a function of $t$ for the entire range of doping ($x = 0.0$-$0.2$) in Fig. 3b. The matching of the patterns of $T_{OO}$-$t$ with those of $D$, $b$, and <Mn-O> versus $t$ is remarkably well. It shows that in the low doping level ($x < 0.1$), the 'intrinsic' distortion does not really maximize *within the range of t accessed here* while it maximizes and influences the $T_{OO}$ significantly beyond $x \sim 0.1$. The plot of $\Delta T_{OO}/\Delta t$ versus $x$ (Fig. 4), in addition, shows that with the increase in doping level ($x$), the slope $\Delta T_{OO}/\Delta t$ first decreases and then branches out between positive and negative values reflecting the non-monotonic $T_{OO}$-$t$ pattern in the doping regime $0.1<x\leq0.2$. This plot highlights the fact that the rate of variation of $T_{OO}$ with $t$ comes down with the increase in $x$ and the non-monotonicity in $T_{OO}$-$t$ sets in only when the rate has dropped from maximum. In other words, maximization of 'intrinsic' octahedral-site distortion can be observed in a region where dependence of $T_{OO}$ on $t$ has weakened down to a certain extent. Of course, if we could extend the range of $t$ by carrying out the measurements across $La_{1-x}A_xMnO_3$ to $Lu_{1-x}A_xMnO_3$, we could, possibly, have observed onset of the branching at a lower tolerance factor. However, the qualitative features – drop in the $\Delta T_{OO}/\Delta t$ from maximum with the increase in $x$ followed by onset of branching at a higher $x$ – are expected to remain same even across the entire $t$-space.

The shift in $t_C(x)$ toward higher value with the increase in $x$ shows that the presence of charge carriers and hence emergence of double exchange mechanism induces a qualitative change in the 'intrinsic' octahedral-site distortion itself. It is no longer maximizing at a universal $t_U$ or $<r_R>_U$. Instead, with the increase in doping it is continuously shifting towards higher tolerance factor reflecting the fact that the emergence of anomalous phase in between orthorhombic and cubic phases with the increase $t$ is taking place at a higher and higher tolerance factor. The orthorhombic phase region in the entire $t$-space is progressively extended with the increase in doping.

The maximization of $D$, $b$, <Mn-O>, and $T_{OO}$ at an intermediate $t_C$ or $<r_R>$ possibly signifies dominance of the role of electron-lattice interaction resulting from Jahn-Teller effect and R-O bond covalency in governing the orbital order. While in the case of undoped $RMnO_3$, $RVO_3$, $RTiO_3$ etc, it appears that the strength of the contribution of electron-lattice interaction in governing the orbital order and $T_{OO}$ is system independent – both distortion and $T_{OO}$ maximize at a universal $<r_R>_U$ ~1.110 Å for all the systems – in quite contrast, in doped systems one observes a doping-dependent shift in the maximization toward higher tolerance factor. This observation points out that the electron-lattice interaction in doped manganites maximizes at a higher and higher tolerance factor i.e., in systems where overall distortion is less.

In summary, we show that both 'intrinsic' octahedral-site distortion and $T_{OO}$ maximize at a 'doping-dependent' critical tolerance factor $t_C(x)$ in doped $R_{1-x}A_xMnO_3$ ($x$ = 0.0-0.2) family reflecting progressive extension of the orthorhombic phase region in the $t$-space and maximization of the strength of electron-lattice interaction in governing the orbital order at a progressively smaller distortion level. These features are in quite contrast to what has been observed in undoped orthorhombic perovskite systems – maximization of $T_{OO}$ and 'intrinsic' distortion at a universal $<r_R>_U$. We also show that the impact of 'intrinsic' distortion on $T_{OO}$ becomes apparent only when the dependence of $T_{OO}$ on $t$ has weakened.

We thank P. Mandal (SINP, Kolkata, India) and S. Elizabeth (IISc, Bangalore, India) for the single crystals. One of the authors (PM) acknowledges support from CSIR, Government of India, during this work.

*e-mail: dipten@cgcri.res.in

[1] See, for example, K.I. Kugel and D.I. Khomskii, Sov. Phys. Usp. **25**, 231 (1982).

[2] C. Lin and A.J. Millis, Phys. Rev. B **78**, 174419 (2008); S. Okamoto, S. Ishihara, and S. Maekawa, Phys. Rev. B **65**, 144403 (2002).

[3] T. Mizokawa, D.I. Khomskii, and G.A. Sawatzy, Phys. Rev. B **60**, 7309 (1999).

[4] E. Pavarini and E. Koch, arXiv:cond-mat/0904.4603v1 (2009).

[5] See, for example, D. Bhattacharya, P.S. Devi, and H.S. Maiti, Phys. Rev. B **70**, 184415 (2004); see also, G. Maris, V. Volotchaev, and T.T.M. Palstra, New J. Phys. **6**, 153 (2004).

[6] J.-S. Zhou and J.B. Goodenough, Phys. Rev. B **77**, 132104 (2008); *ibid.*, Phys. Rev. Lett. **96**, 247202 (2006); *ibid.*, Phys. Rev. Lett. **94**, 065501 (2005).

[7] S. Miyasaka, Y. Okimoto, M. Iwama, and Y. Tokura, Phys. Rev. B **68**, 100406(R) (2003).

[8] P. Mondal, D. Bhattacharya, and P. Choudhury, J. Phys.: Condens. Matter **18**, 6869 (2006).

[9] P. Mondal, D. Bhattacharya, P. Choudhury, and P. Mandal, Phys. Rev. B **76**, 172403 (2007).

[10] M.-H. Sage, PhD Thesis (University of Groningen, Netherlands, 2006).

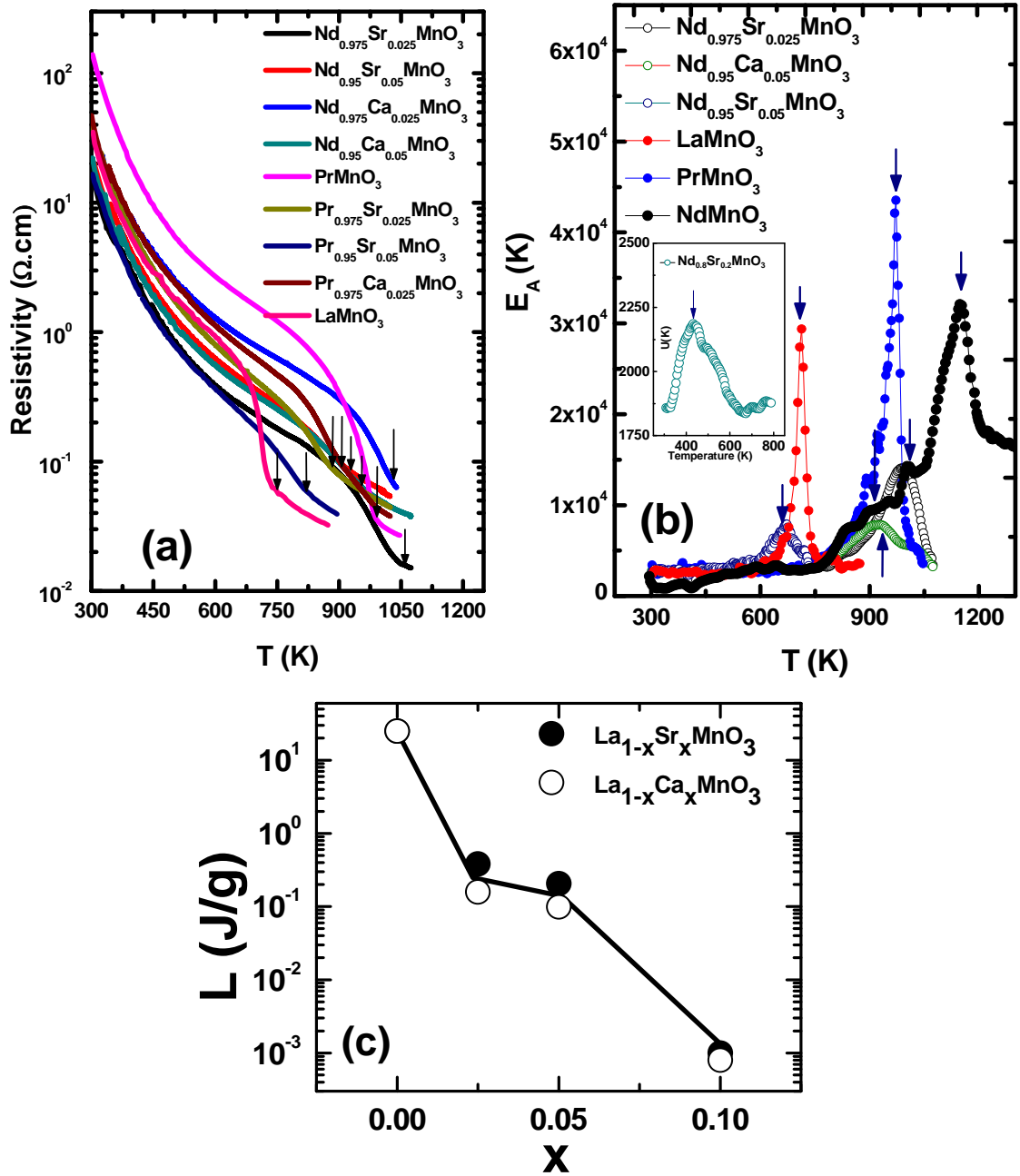

**Fig. 1.** (a) The resistivity ($\rho$) versus temperature ($T$) plots for a few representative doped manganite systems; the arrows mark the $T_{OO}$; (b) $d\ln(\rho/T)/d(1/T)$ versus temperature plots; the width of the peak signifies the width of the transition zone; the arrows mark the $T_{OO}$; (c) the latent heat of orbital order-disorder transition versus doping level ($x$); the latent heat is calculated from the area under the peak of a DSC thermogram; the latent heat decreases fast with the increase in doping signifying a crossover from first to higher order transition with the increase in charge carrier concentration.

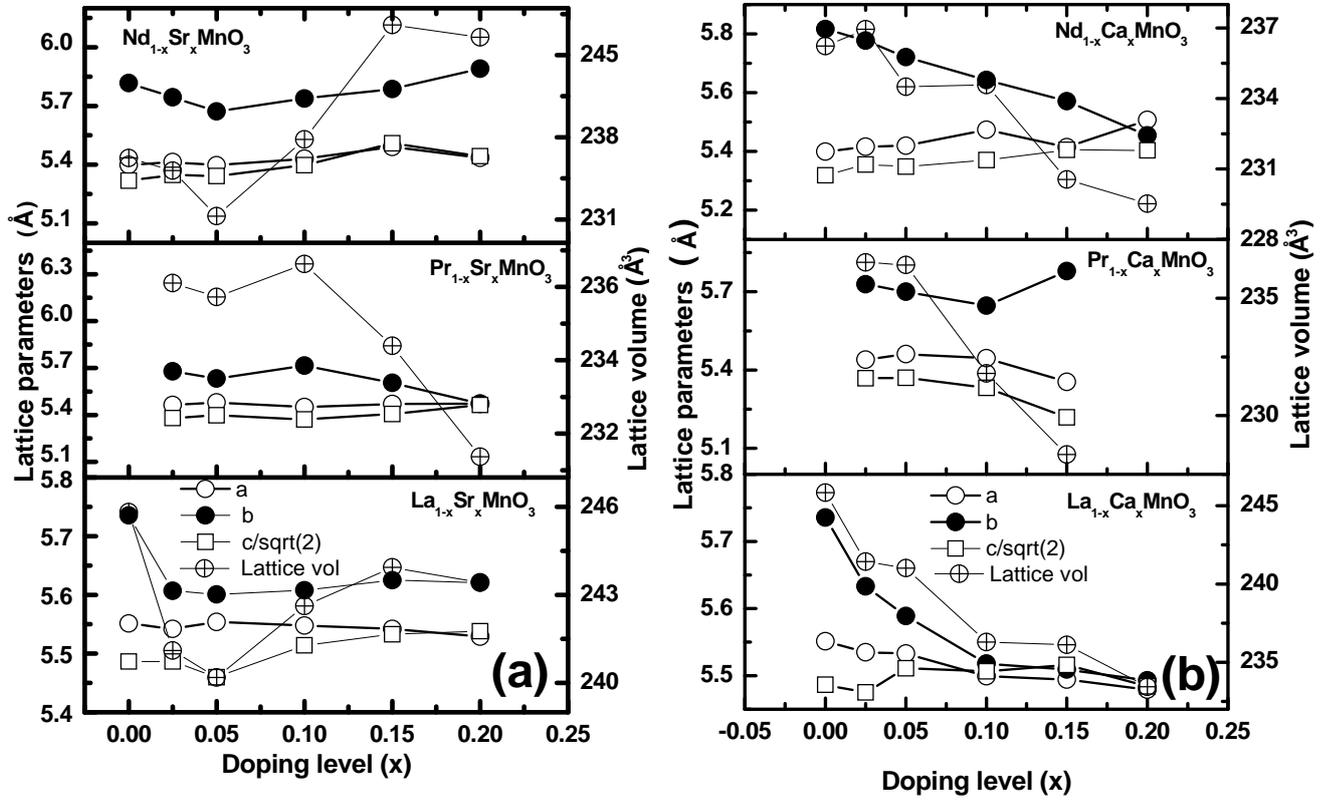

**Fig. 2.** The lattice parameters $a$, $b$, $c/\sqrt{2}$, and lattice volume $V$ are plotted as a function of doping level ($x$) for the (a) Sr-doped systems and (b) Ca-doped systems; while $V$ decreases initially and then rises with the increase in $x$ for Sr-doped systems, it registers a monotonic decrease for the entire range of $x$ in the case of Ca-doped systems.

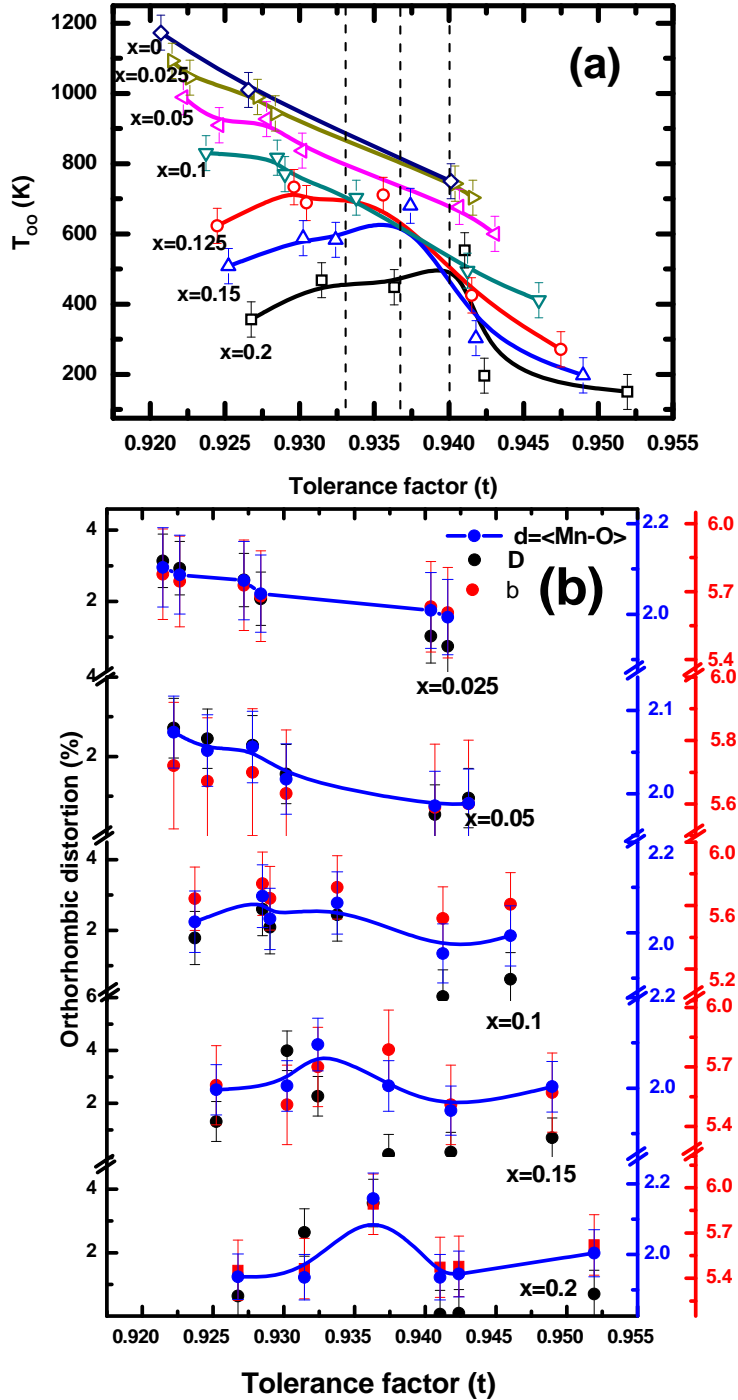

**Fig. 3.** (a) $T_{OO}$ versus tolerance factor ($t$) plots for doped manganites with different doping levels ($x$) are shown; within the range of $t$ accessed here, the plots exhibit a switch from monotonic to non-monotonic pattern reflecting influence of 'intrinsic' octahedral-site distortion; (b) $D$, $b$, and $d$ = <Mn-O> bond length are plotted as a function of $t$; the switch from monotonic to non-monotonic pattern with the increase in x appears similar to the switch in $T_{OO}$-$t$ plots.

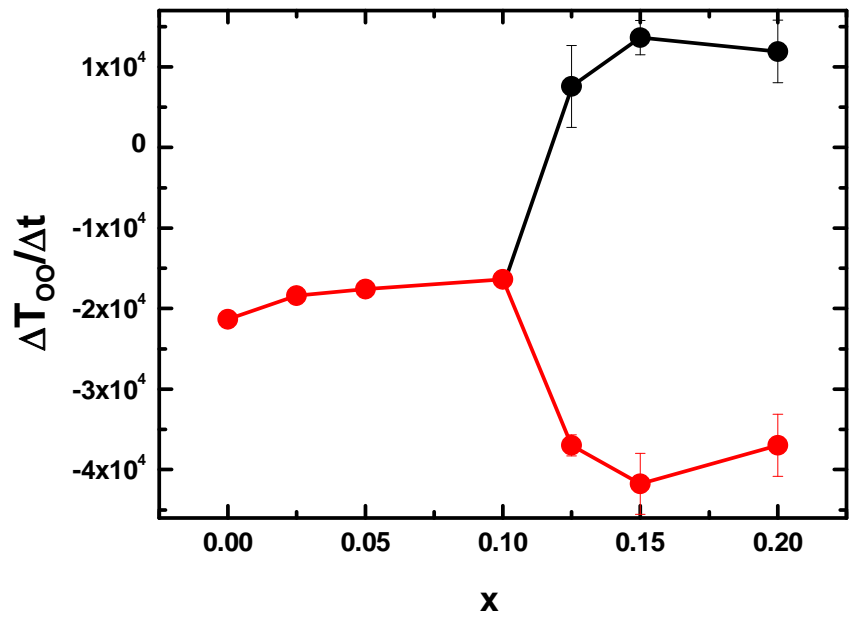

**Fig. 4.** The dependence of $T_{OO}$ on $t$ as a function of doping level ($x$) in doped manganites; the dependence weakens slightly and then branches out reflecting the non-monotonic pattern in the $T_{OO}$-$t$ plots within the $t$ range accessed in this work.